\documentclass{aastex61}

\newcommand{\degree}{^{\circ}}

\received{Jan 1, 2000}
\revised{Jan 1, 2000}
\accepted{Jan 1 2000}
\submitjournal{AJ}

\shorttitle{Planetary meteoroid environments}
\shortauthors{Wiegert et al.}
\begin{document}

\title{Measuring the meteoroid environments of the planets with meteor detectors on Earth}

\

\correspondingauthor{Paul Wiegert}
\email{pwiegert@uwo.ca}

\author[0000-0002-1914-5352]{Paul Wiegert}
\affiliation{Dept. of Physics and Astronomy, The University of Western Ontario London Canada}
\affiliation{Centre for Planetary Science and Exploration (CPSX), The University of Western Ontario London Canada}

\author{Peter Brown}
\affiliation{Dept. of Physics and Astronomy, The University of Western Ontario London Canada}
\affiliation{Centre for Planetary Science and Exploration (CPSX), The University of Western Ontario London Canada}

\author{Petr Pokorny}
\affiliation{Dept. of Physics, The Catholic University of America, Washington D.C., USA}

\author{Karina Lenartowicz}
\affiliation{Dept. of Physics and Astronomy, The University of Western Ontario London Canada}

\author{Zbyszek Krzeminski}
\affiliation{Dept. of Physics and Astronomy, The University of Western Ontario London Canada}



\begin{abstract}

We describe how meteors recorded at the Earth can be used to partly reconstruct the meteoroid environments of the planets if a large sample (i.e. millions of orbits at a minimum) is available. The process involves selecting from the Earth-based sample those meteors which passed near a planet's orbit prior to arriving at the Earth, and so carry information about the planetary meteoroid environment. Indeed this process can be extended to any location in the Solar System, though some regions of space are better sampled than others.

From such a reconstruction performed with data from the Canadian Meteor Orbit Radar (CMOR), we reveal that Mars has apex, helion, anti-helion and toroidal sporadic sources, much as Earth does.  Such reconstructions, albeit partial, have the potential to provide a wealth of detail about planetary meteoroid environments, and to allow for the ground-truthing of model meteoroid populations without {\it in situ} sampling.

\end{abstract}




\section{Introduction}

The meteoroid environment at the Earth has been measured over a large particle size range
 by a number of techniques, while the environments of the
other planets are much less well known. The difference is largely due
to the fact that the equipment used to measure meteors at Earth, such
as camera networks and meteor radars, which use the atmosphere as a detector cannot easily be transported and
deployed on other planets by spacecraft.

Some dust measurements have been performed of the meteoroid environment
away from the Earth, \citep{Staubach2002}. However, these are primarily in-situ measurements by spacecraft
 impact detectors and therefore limited to small
collecting areas. As recent examples, the Cosmic Dust Detector on the
Cassini spacecraft at Saturn \citep{sraahralt04, kemsraalt04}, and the Student Dust Counter on the New Horizons mission to Pluto \citep{horhoxjam08,popjamjac10,szapiqhor13} both have sensitive areas of only 0.1~m$^2$. They are therefore unable to sample particles much larger than a few microns. In contrast, ground-based meteor
detectors might monitor thousands of square km of sky or more
e.g. $\sim 10^6$ km$^2$ for the European Fireball Network
\citep{obemolhei98}. 

Several works have predicted the existence of meteor showers on other planets based on known comet/meteoroid stream orbits (eg. \citep{Ch2010},\citep{Selsis2004}) with close nodal points to other planetary orbits. However, observations of planetary meteoroid environments other than Earth at sizes larger than a few microns are limited to a short optical survey at Mars \citep{Domokos2007}, individual impacts in the Jovian atmosphere \citep{Hueso2013} and indirect inference of meteoroid impacts on Mercury based on exospheric gas measurements \citep{Christou2015}.    

Here we report on a technique that allows us to  study the meteoroid
environment at a planet in our solar system directly from Earth. This works for any
solar system planet, and indeed arbitrary locations in interplanetary space. It uses the same equipment as Earth-based meteor studies,
whether camera, radar or otherwise. It simply involves selecting from
the observed terrestrial meteoroid orbit sample those particles which passed
near the orbit of another planet prior to arriving at Earth. Particles which
pass near a planet before arriving at Earth are {\it de facto} part of
that planet's meteoroid environment, and recognizing them as such, we
can partially reconstruct the planetary meteoroid environment (PME)
through inversion of these carefully chosen meteoroid orbits as observed at Earth. An important obstacle for this technique is the small number of meteoroids
travelling from the vicinity of any given planet to the Earth. As a result, a necessary condition for this technique to be useful is
a large sample of measured meteoroid orbits . We can improve our sampling if we
are willing to assume that the meteoroid environment of the planets is
approximately azimuthally symmetric about its orbit: this is known to
be approximately true for the Earth \citep{camjon06}. In this case,
the sample of meteoroids measured at Earth which pass sufficiently near
the {\it orbit} of a given planet (rather than those which pass
sufficient near the planet itself) provides a proxy for the time
averaged PME.  Given sufficient numbers of meteoroid orbits, we will show that
such a sub-sampling provides useful information on PMEs, and show that
from it we can predict  (as just one example) that features in the spatial radiant distribution
at Mars are reminiscent of those seen in the sporadic distribution at
Earth.

This approach is not without its limitations: geometrical restrictions
arise from the need for the meteoroid orbit to pass near both the
planet in question and the Earth. We also need a $very$ large sample
of meteors to get usable statistics. Here the Canadian Meteor Orbit
Radar (CMOR, \cite{webbrojon04}) which measured over 4.3 million
meteors orbits in the interval 2011-2014 is the sample that will be
used. Even with this vast data set, only a small fraction ($\sim~1\%$)
are useable for our purposes: a smaller data set will scarcely
do. Nonetheless, partial data is better than none, and provides a
first look at what otherwise would remain undetected entirely.

% From the files that I got from Petr, there were a total of 4,382,353 meteors from 2011-2014
% Not sure what fraction of these you are interested in highlighting 

\section{Method}

Our intent here is to provide a proof of concept and present an
initial reconnaissance of what portions of the planetary meteoroid
environments can be sampled from Earth, and what they look like in
their broad strokes.  The key to this technique is determining which of
the CMOR meteoroid orbits  passed near the orbit of another planet prior to its
detection at Earth. The CMOR meteor patrol radar has been described
elsewhere \citep{webbrojon04}. Here we will use the data measured from  CMOR
 since 2011, as this is the time period after which a total of six-stations
 began operation \citep{browerwon12} from an original network of three and hence the precision of orbits is significantly better on the average since 2011. To determine whether a given
meteoroid orbit is part of another planet's PME, we compute the Minimum Orbital
Intersection Distance (MOID) between its pre-atmospheric orbit and the
planet's osculating orbit.  If the MOID is less than a particular value (which varies by
planet), we will consider it part of our PME sample. The values chosen
here are of the same order as the planetary Hill spheres: 0.01~AU for
the terrestrial planets, and 1~AU for the outer planets.  For this initial
study, we will ignore the effects of instrumental error on the MOID
calculation, though inclusion of  uncertainties could be used to
refine the calculated PME.

The sample of CMOR meteoroids that passed near a planet's orbit prior to its
arrival at Earth gives us a first look at the planetary meteoroid
environment. To obtain a more quantitative result, we have to consider
the efficiency with which particles can travel from the planets to
us. There are at least two parts to this calculation.

The first relates to the fraction of meteoroid orbits that have a
geometry suitable to reach the Earth; the second has to do with the
'field of view' of the radar itself, which cannot observe all
radiants which reach our planet. These biases will be addressed in the next
two subsections.

\subsection{Meteoroid orbital transfer efficiency from the planets to Earth}
\label{te:transferefficiency}

Some orbits passing near the planet's orbit can reach the Earth, others cannot
on purely geometrical grounds.  To examine this, we analyse a random
sample of hypothetical meteoroids passing near each of the planets to
determine which can reach the Earth afterwards. This work is a simple
extension of the scattering work done as part of \cite{wie14}, and
allows an assessment of how much of a PME we can hope to sample under
ideal conditions.

We will use the planet Mars as an example here for concreteness. If we
imagine all possible radiant directions and speeds on Mars' meteor sky,
some would (if they weren't destroyed by their encounter with Mars) go
on to reach the Earth. We can compute numerically which of these
radiants could have been sampled at Earth, to get a picture of the
portion of Mars' meteor sky we can in principle sample. We refer to the function defining whether or not a given radiant at  Mars can reach Earth
as the Transfer Function (TF). 

\begin{figure}[ht!]
    \centering
    %\vspace*{-1cm}
    \includegraphics[width=25pc]{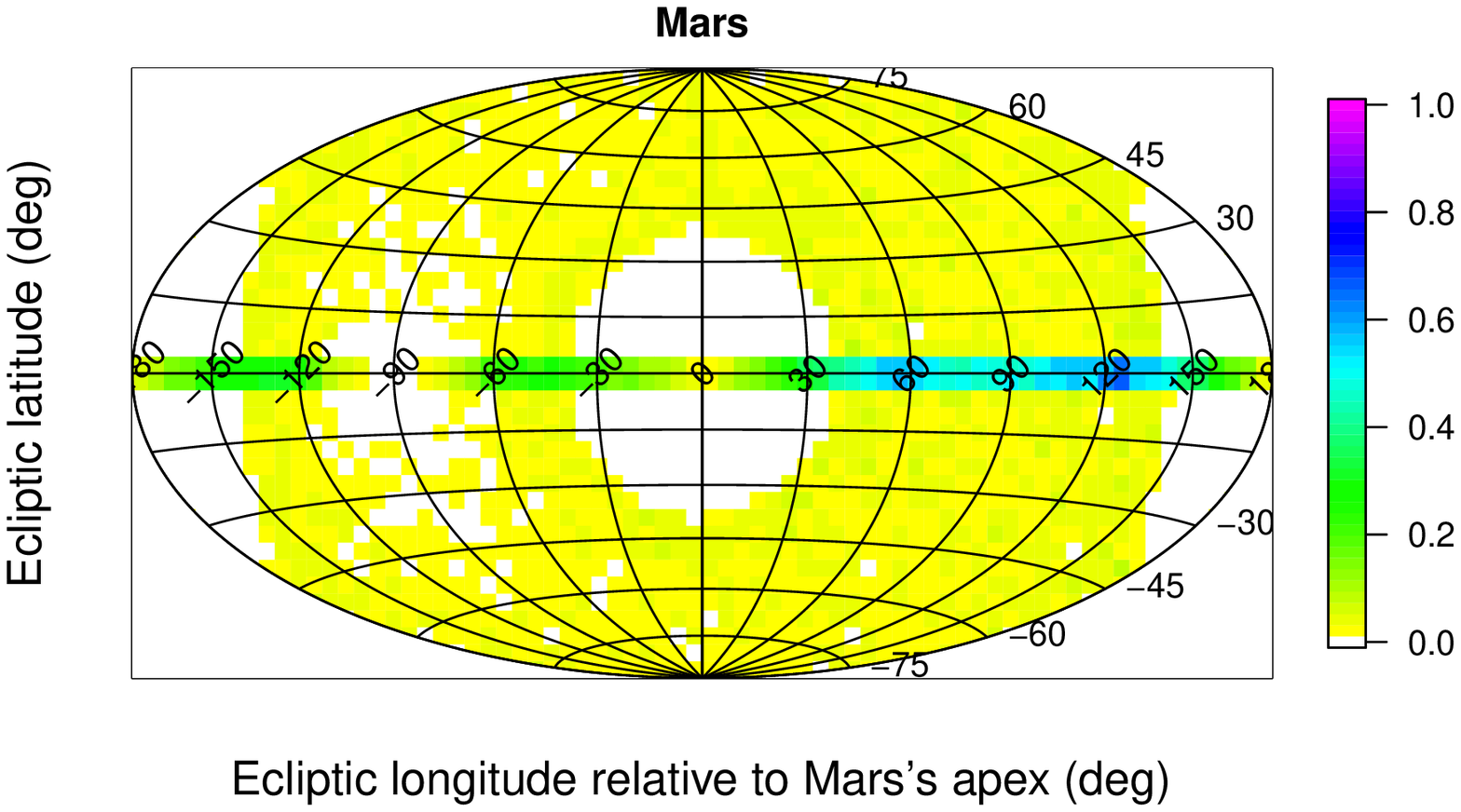}
    %\vspace*{-1cm}
    %\hspace*{-0.7in}
    \begin{tabular}{ccc} % textwidth means that the image will be the width of the text
        \includegraphics[width=12pc]{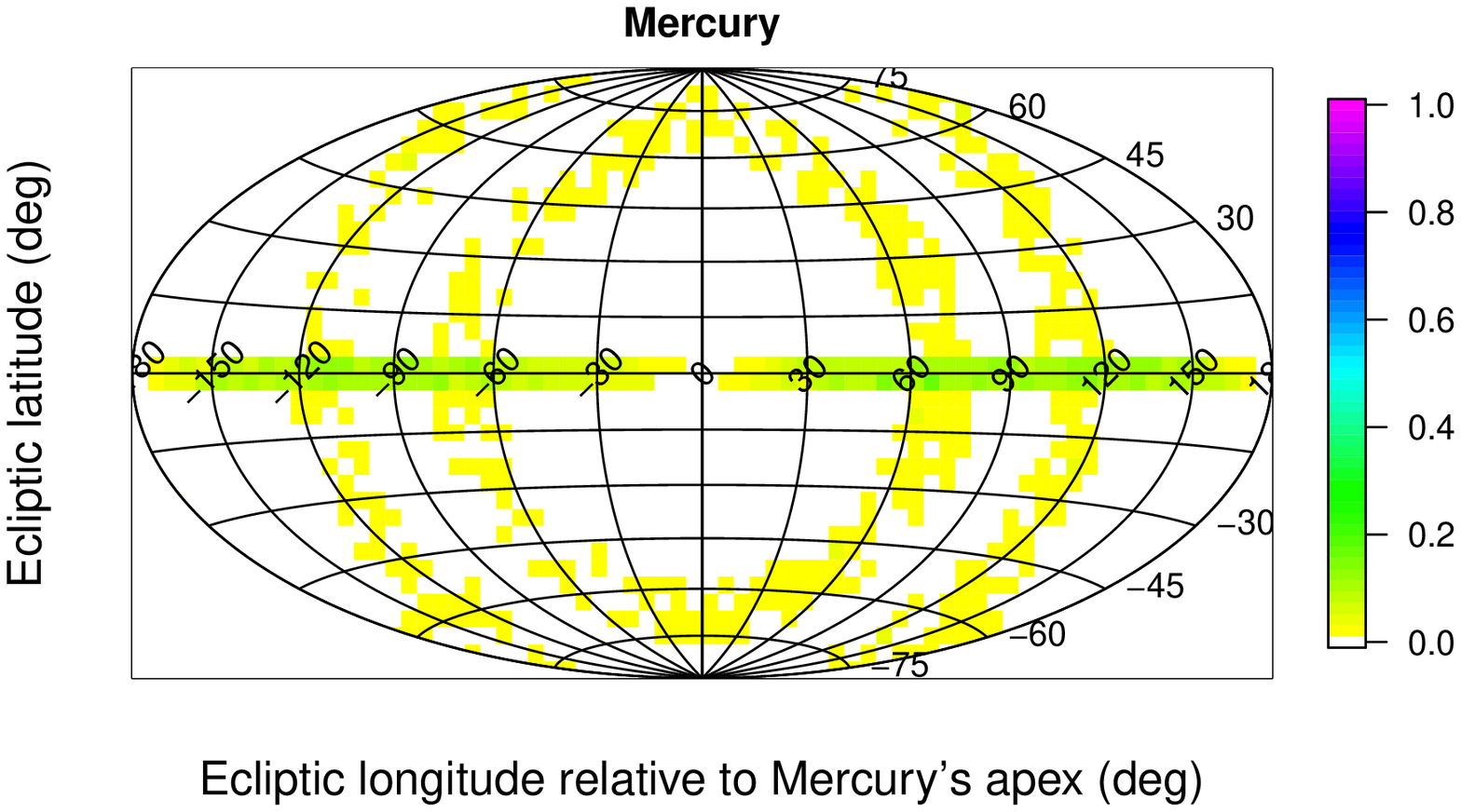} &
        \includegraphics[width=12pc]{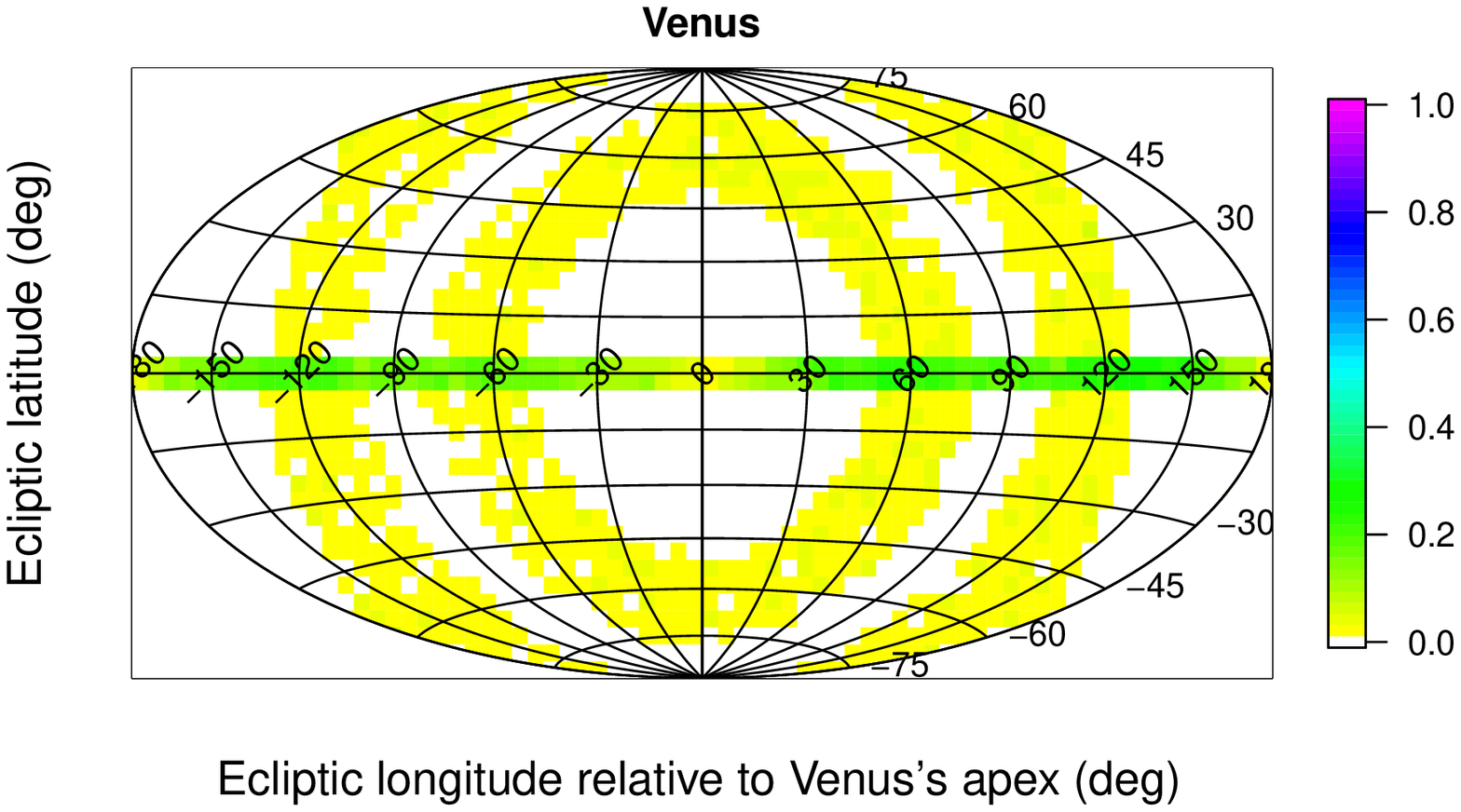} &
        \includegraphics[width=12pc]{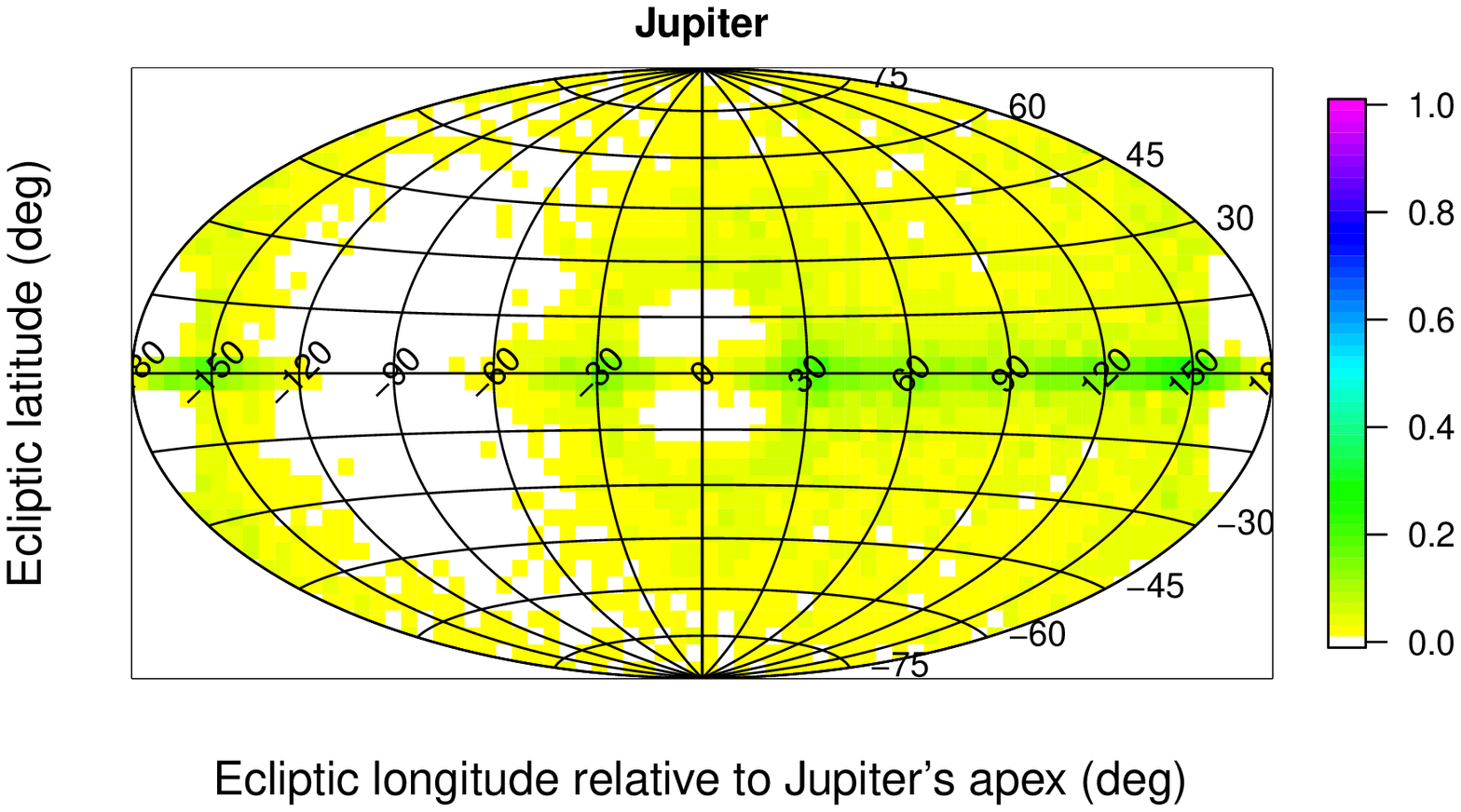} \\
        \includegraphics[width=12pc]{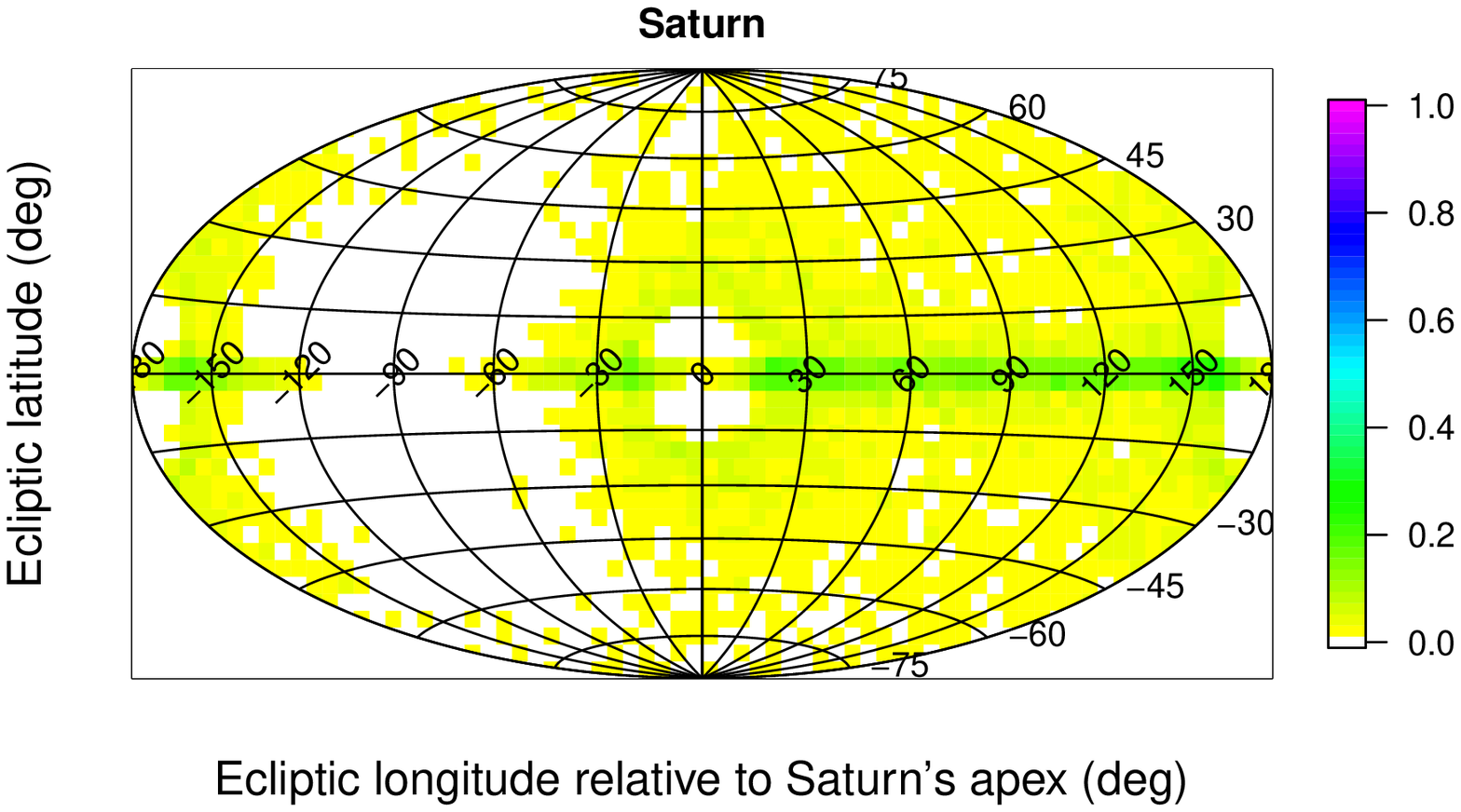} &
        \includegraphics[width=12pc]{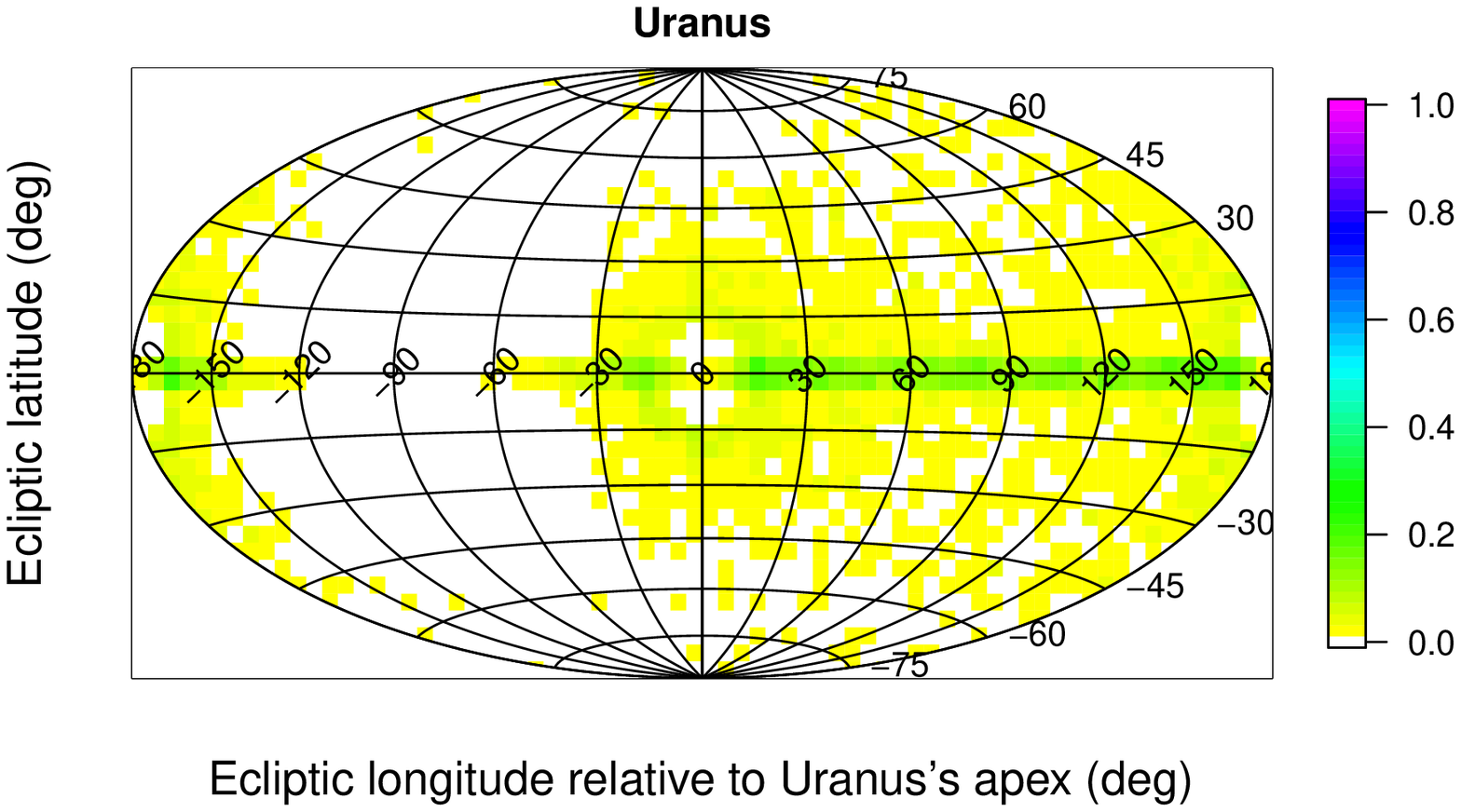} &
        \includegraphics[width=12pc]{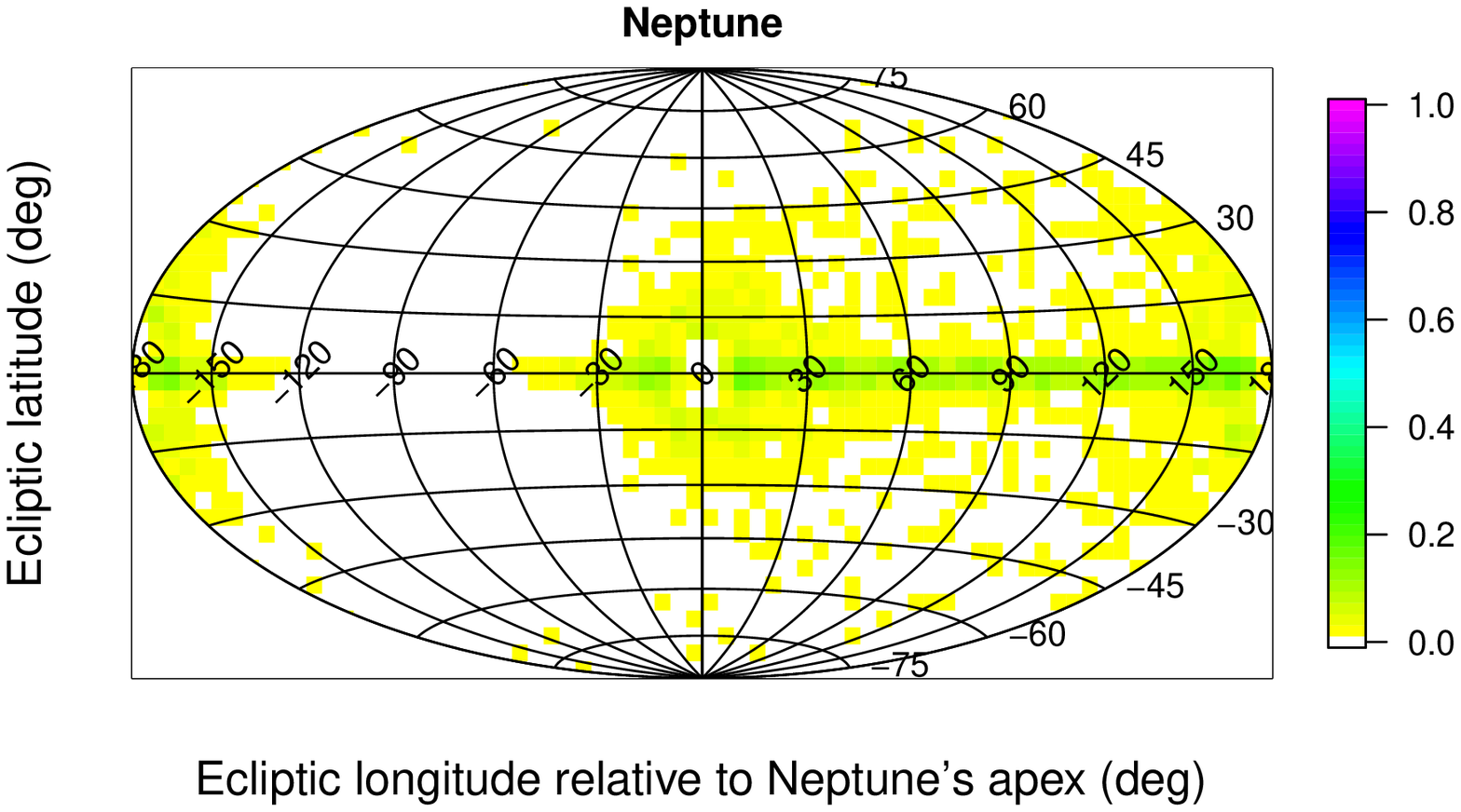} \\
        %\multicolumn{2}{c}{\includegraphics[width=0.4\paperwidth]{figs/TFunction/Neptune_t.eps}}
    \end{tabular}
    
    \caption{The meteor sky at each planet colour-coded by the fraction of
  radiants which can reach the Earth, in apex-centered coordinates.
  The apex of the planet's way is in the center; the Sun is at $-90\degree$
  longitude, and opposition at $+90\degree$. White regions indicate
  radiants for which less than 0.1\% of the possible speeds can reach Earth. \label{fi:transferfunctions}}

\end{figure}

To compute the TF, one million hypothetical meteoroids at Mars are
created with radiants that are drawn randomly on Mars' sky, and for
each a speed is randomly drawn from a uniform distribution between
zero and the heliocentric escape speed at Mars. The fraction of these
that could reach Earth are counted. The criterion for reaching Earth
is that the meteoroid orbit have a node between 0.9 and 1.1 AU from
the Sun or that its heliocentric inclination be less than $3\degree$ or more than $177\degree$. The inclination choice used is based on the size of the Earth's Hill sphere, 0.01~AU, which subtends an angle of $3\degree$ when seen from the Sun. The addition of the inclination criterion is a first order attempt to account for the fact that meteors with orbits very nearly in the plane of Earth's orbit can strike our planet even if they do not have a node near 1~AU; a more sophisticated criterion based on the MOID could be used, but is unlikely to affect the results significantly. A plot of the TF for Mars, that is the meteor radiants at Mars which can reach Earth, is in the top panel of  Fig.~\ref{fi:transferfunctions}.

Each pixel in the top panel of Fig.~\ref{fi:transferfunctions} represents the fraction of meteors
that arrive at Mars from a particular radiant, and which subsequently
can reach Earth.  Note that the radiants in Fig.~\ref{fi:transferfunctions} are
not the meteor's radiants at Earth (which are different) but those they
would have had at the planet itself had they struck it.

Any regions in white in Fig.~\ref{fi:transferfunctions} represent 'forbidden
regions' in that no meteors at any speed considered here can travel
from the planet to the Earth, and so we are inherently blind to these radiant zones.
The meteoroids at Mars which have Martian radiants in the ecliptic plane are the best sampled at
Earth, however, there are numerous radiants at higher ecliptic
latitudes that are also accessible. If Mars has six sporadic meteor
sources similar to those at Earth \citep{jonbro93},
The top panel of Fig.~\ref{fi:transferfunctions} indicates that we can expect to sample all of them
at Earth, though many only at a very low level ($\sim
5$\%). Nonetheless, this provides a first important result from this
work: meteor detectors on Earth have access to a fraction of the
meteoroid environment at Mars and in fact, at the other planets as
well, as will be described later.

Our transfer efficiency is idealized for simplicity. It assumes that
both the Earth and Mars are on circular orbits, that the meteoroid
was not appreciably deflected by the planet when it passed it, and
that the intersection criterion described above is applicable. Though
reasonable approximations, they are not perfect and as a result we
will occasionally find that CMOR meteors are detected from these
forbidden regions, but these discrepancies do not affect our broader
argument.

The TFs for other planets are also shown in the lower panels of Figure~\ref{fi:transferfunctions}.
The fraction of meteors which can reach Earth from each of the
planet's PME is in Table~1 and ranges from 9.0\% at Mars
down to 2.8\% for Mercury. Thus a wide range of radiants across a
planet's meteor sky are in principle accessible from Earth. As a rule, the ecliptic
plane is always best sampled, as the geometrical intersection constraint is
easier to satisfy. The apex region (near the zero of the apex-centered
ecliptic latitude and longitude) is often completely inaccessible, but
much of the rest of the planet's meteor sky is sampled at some
level. 

The idealized TF described above allows us to assess how much of the available phase space can be measured from Earth. Also interesting is the fraction of the $populated$ phase space at each planet which can reach Earth. This is, of course, unknown at this time but we can use a model of each planet's meteoroid environment as a proxy. Here we use the planetary meteoroid environment model of \cite{wievaucam09}, a physics-based self-consistent model of the near-Earth meteoroid environment calibrated to CMOR data. Though constructed specifically for Earth, the model contains information on the PME's of all the planets. The terrestrial planets will be reasonably well-represented by this model, but it was not designed for the outer planets. The fraction of each model PME which can be sampled at Earth is in line 2 of Table~1. Here we see that the \cite{wievaucam09} model predictions suggest an even larger fraction of the terrestrial PMEs are accessible to us than was suggested by our uniform model : 9\% for Mercury, 14\% for Mars and 24\% for Venus. The fraction is reduced for the outer planets, to between 1-3\% in all cases.

We conclude that planetary PMEs can be measured at a
significant level from Earth, given a sufficiently large sample of
meteoroid orbits. In particular, these measurements offer the possibility of
'ground-truthing' and calibrating dynamical models of PMEs without
{\it in~situ} measurements.

\subsection{The observational geometry from the CMOR radar}
\label{te:collectingarea}

Having assessed in the previous section that a reconstruction of PMEs is possible in principle, we turn to the observing biases that affect this process. Located in the northern hemisphere, CMOR (43$\degree$16$^\prime$ N,
80$\degree$46$^\prime$ W) can only see meteors arriving from certain
directions. CMOR has a nearly all-sky gain pattern, which means that most radiants which get above the local horizon have some detectability throughout the day. Due to this broad beam pattern, the time-weighted radiant collecting area is purely a function of declination as discusssed in \cite{CB2006}. To determine the effect of this on our sample, we
calculate the relative collecting area $f$ of the radar as a function
of the radiant declination of the arriving meteor. Here we take $f$ to be the time
average over the year of the cosine of the angle between a given
apparent meteor radiant and the local vertical at the radar site, with any
negative values (radiant below the horizon) taken to be zero.  Thus a
 radiant that was
never above CMOR's horizon would have $f=0$. In practice, a radiant at the North equatorial
Pole  has the highest effective value at $\sin(43\degree16^\prime) = 0.69$, a result also found in \cite{CB2006}(see their Fig 6.).
%Figure~\ref{fi:CMORcollectingarea} shows CMOR's relative collecting
%area as a function of ecliptic latitude and longitude.

A weighting of $1/f$ is then applied to that meteor to allow for the
radar's time-integrated collecting area for that radiant. Radiants with $f=0$ are not weighted; they cannot be measured with CMOR but are designated 'inaccessible' (see next section).

There are other observing biases which generically affect back-scatter radar measurements of meteors \citep{GB2004}. For CMOR the most important is the echo height ceiling (which reduces sensitivity to higher speed meteoroids eg. \citep{CB2003}) and the ionization efficiency, which reduces detectability of slower meteoroids \citep{Ye2016}, favoring detection of larger, slower meteoroids. Though these are important for the construction of the best possible PME, here we ignore these effects for simplicity, as our efforts here are towards a proof of concept rather than a high-precision result. Moreover, these effects tend to be size dependent and as we do not use size-frequency distribution in any of our models, as long as CMOR can detect some fraction of the population from any radiant we can use it to probe other PMEs. We also note that we adopt a simple frequentist approach to debiasing the sample here, but Bayesian techniques are likely to perform better when a more complete reconstruction of planetary meteoroid environments is attempted.

\section{Results}
\subsection{Meteoroid environments at the planets} \label{te:meteorenv}

In order to reconstruct the planetary meteoroid environments, each
CMOR meteoroid deemed to have arrived from a PME is assigned a weight
inversely proportional to the relative collecting area of the radar.
No correction for the transfer function is applied because the correct TF to apply depends on the PME in question, which is {\it a priori} unknown.  We do note that the sporadic meteor sources are known to have very broad speed distributions at the Earth \citep{jonbro93}
and so a correction  based on the uniform TF might be correct to zeroth order.  A more accurate approach could involve comparison with a model distribution and/or 'best-fit' approach but this is beyond the scope of this paper: our intent here is simply to provide a first look at what can be recovered about PMEs from Earth-based samples.

\begin{figure}[ht!]
    \centering
    %\vspace*{-2cm}
    \includegraphics[width=20pc]{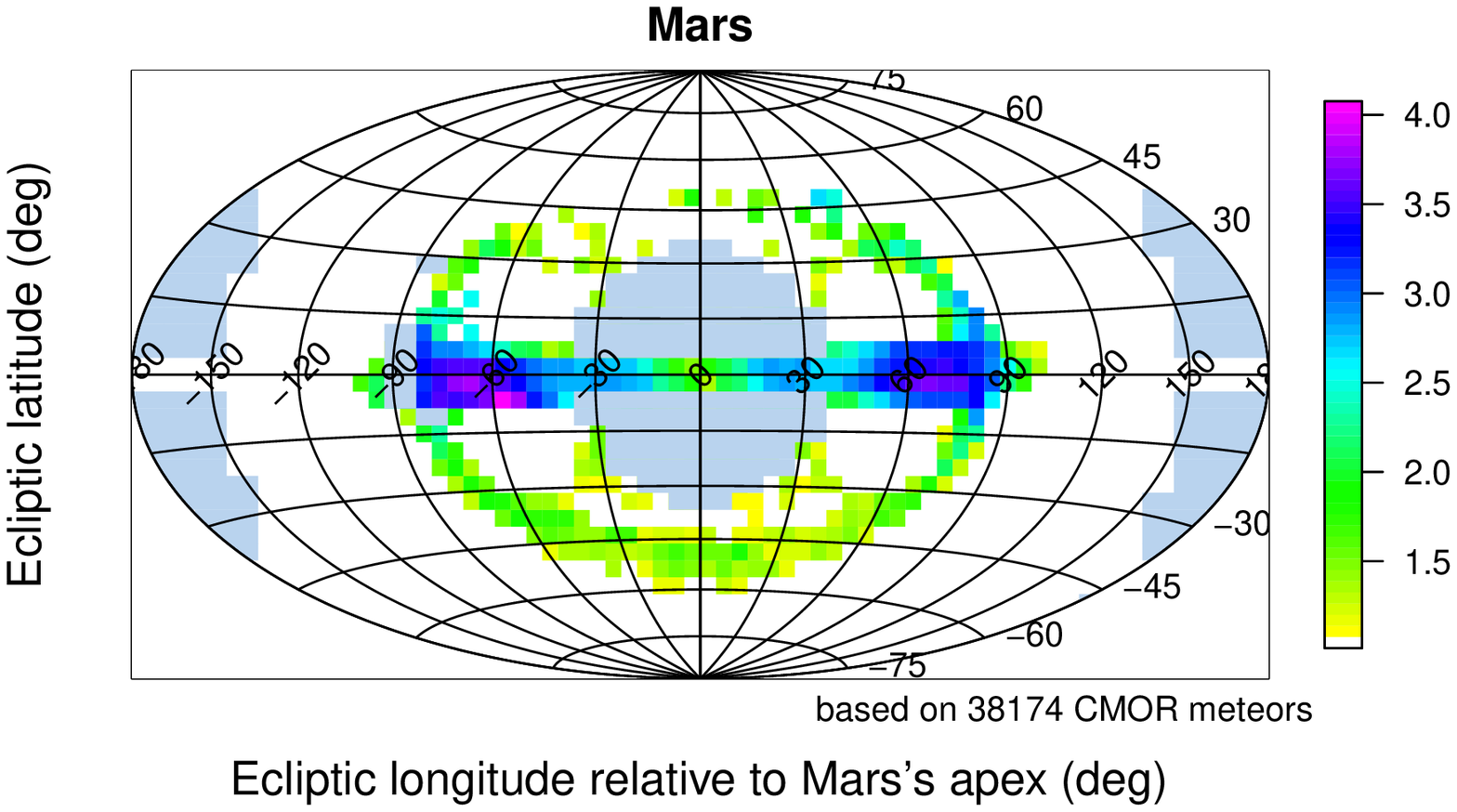}
    %\vspace{-1cm}
    %\hspace*{-0.7in}
    \begin{tabular}{ccc} % textwidth means that the image will be the width of the text
        \includegraphics[width=12pc]{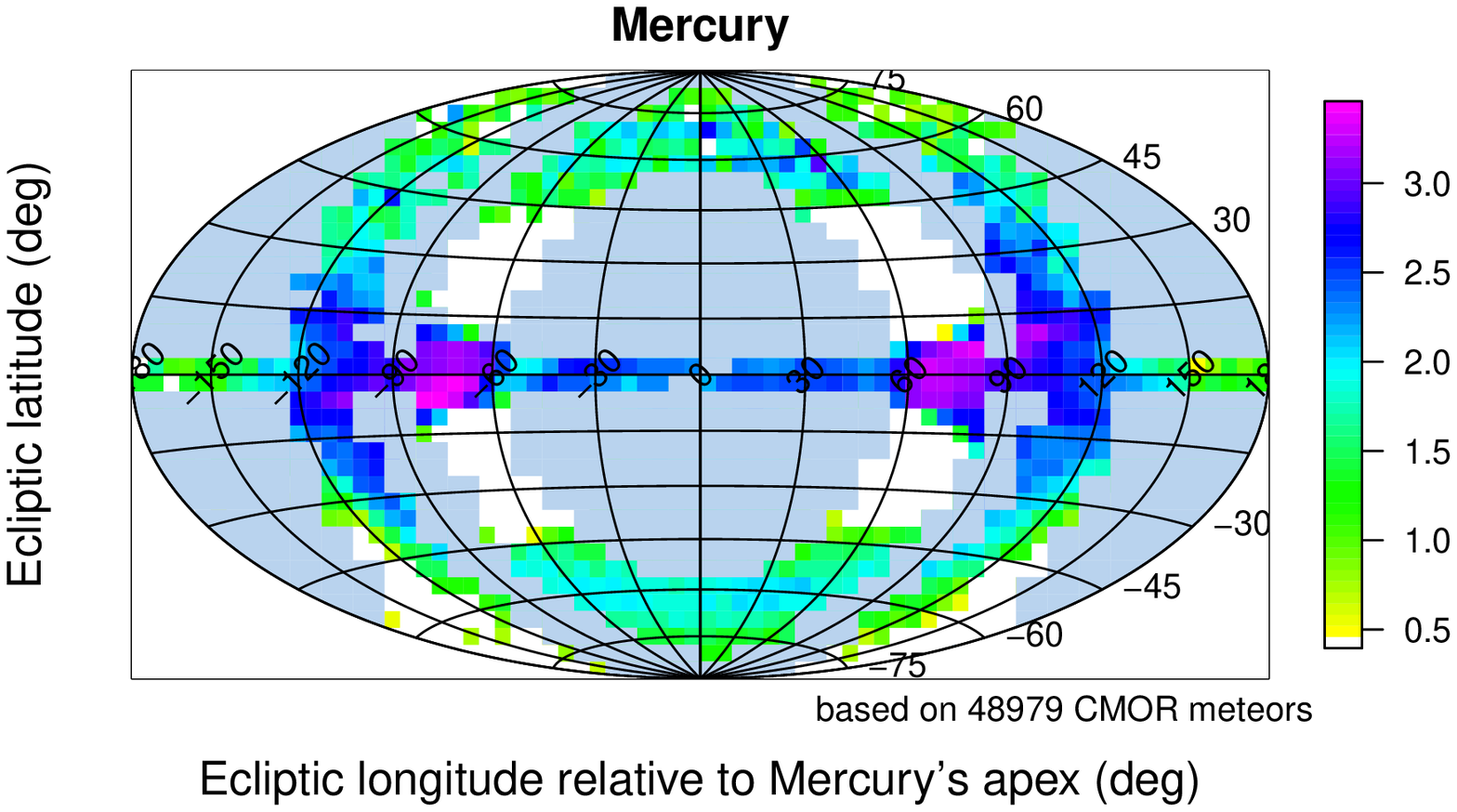} &
        \includegraphics[width=12pc]{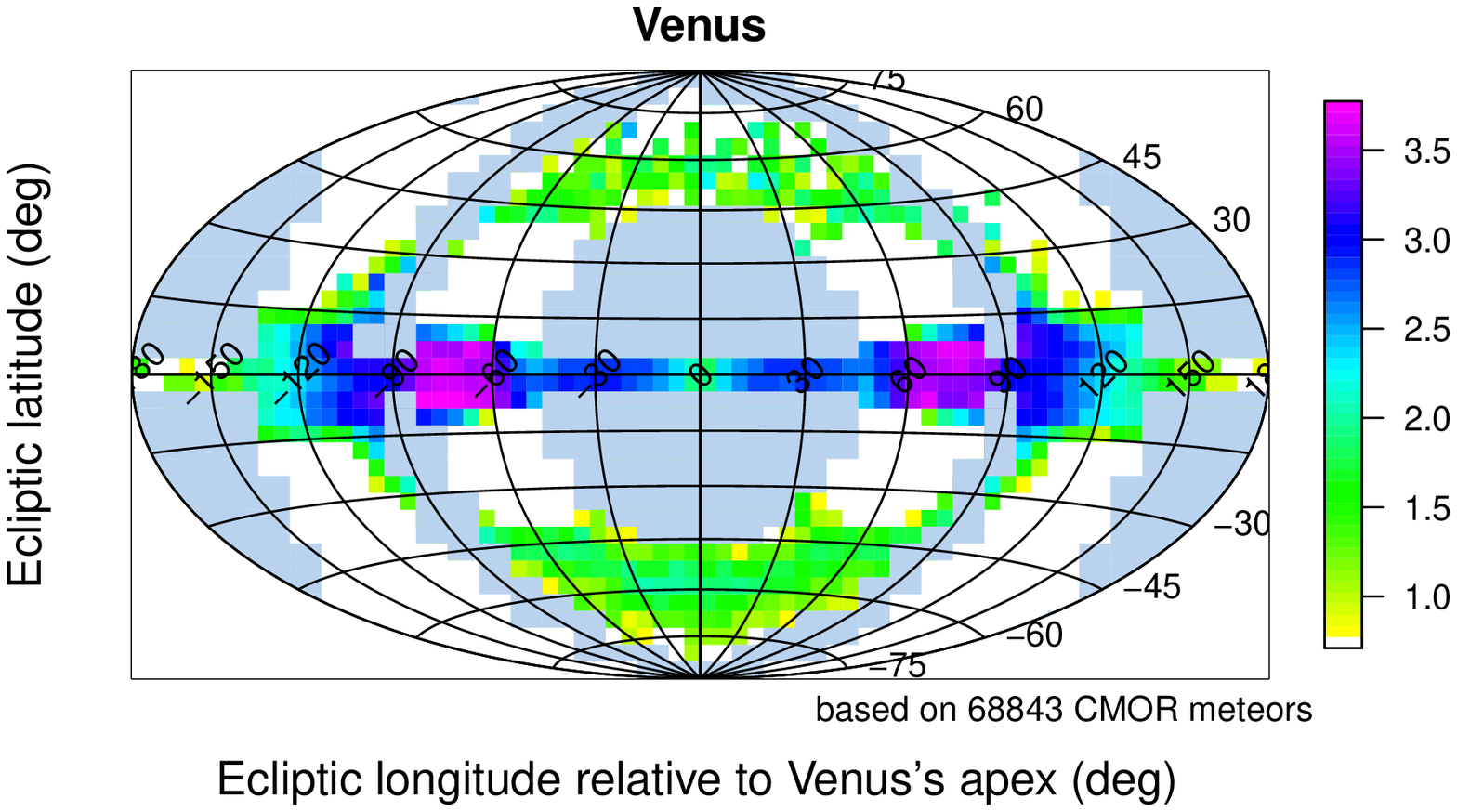} &
        \includegraphics[width=12pc]{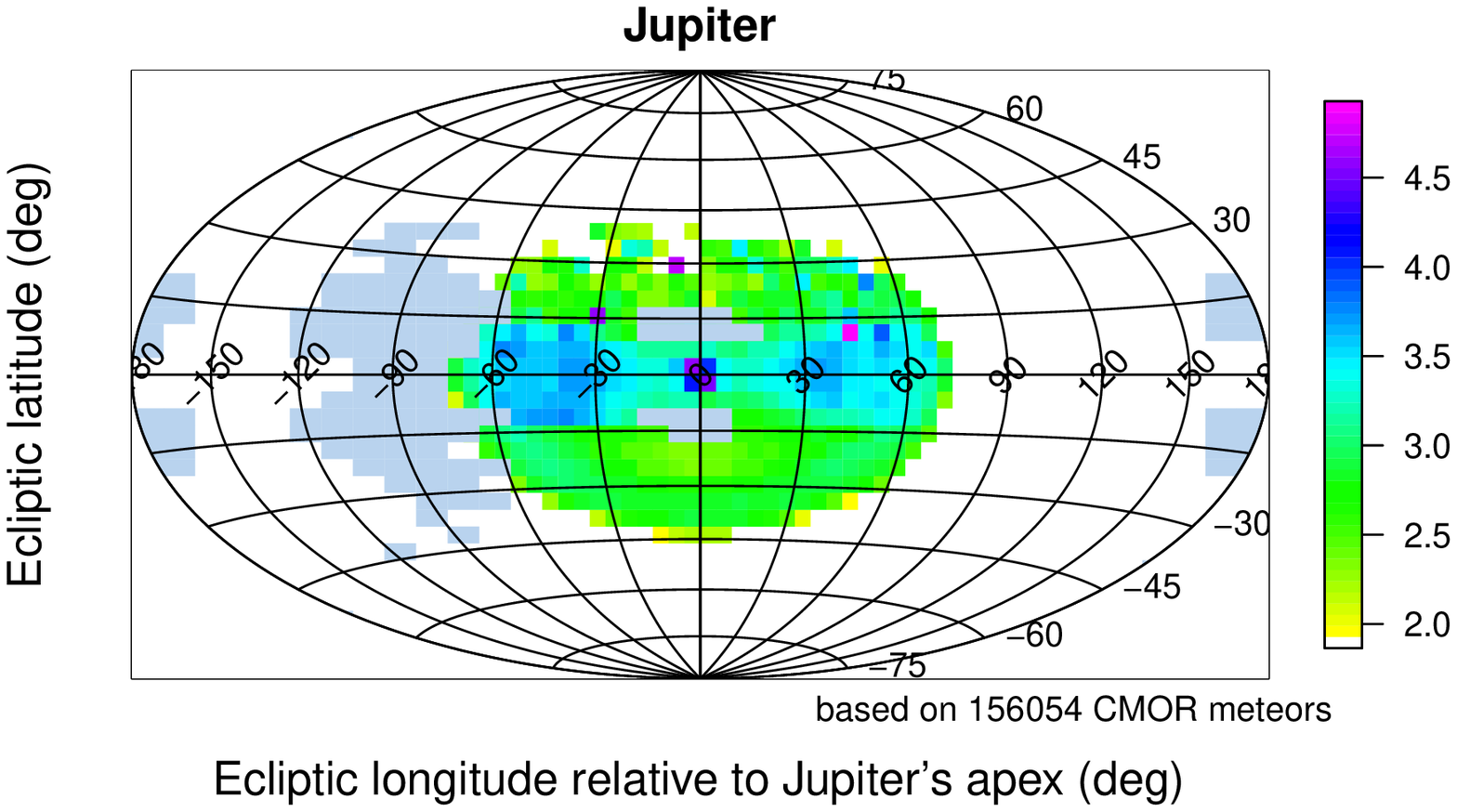} \\
        \includegraphics[width=12pc]{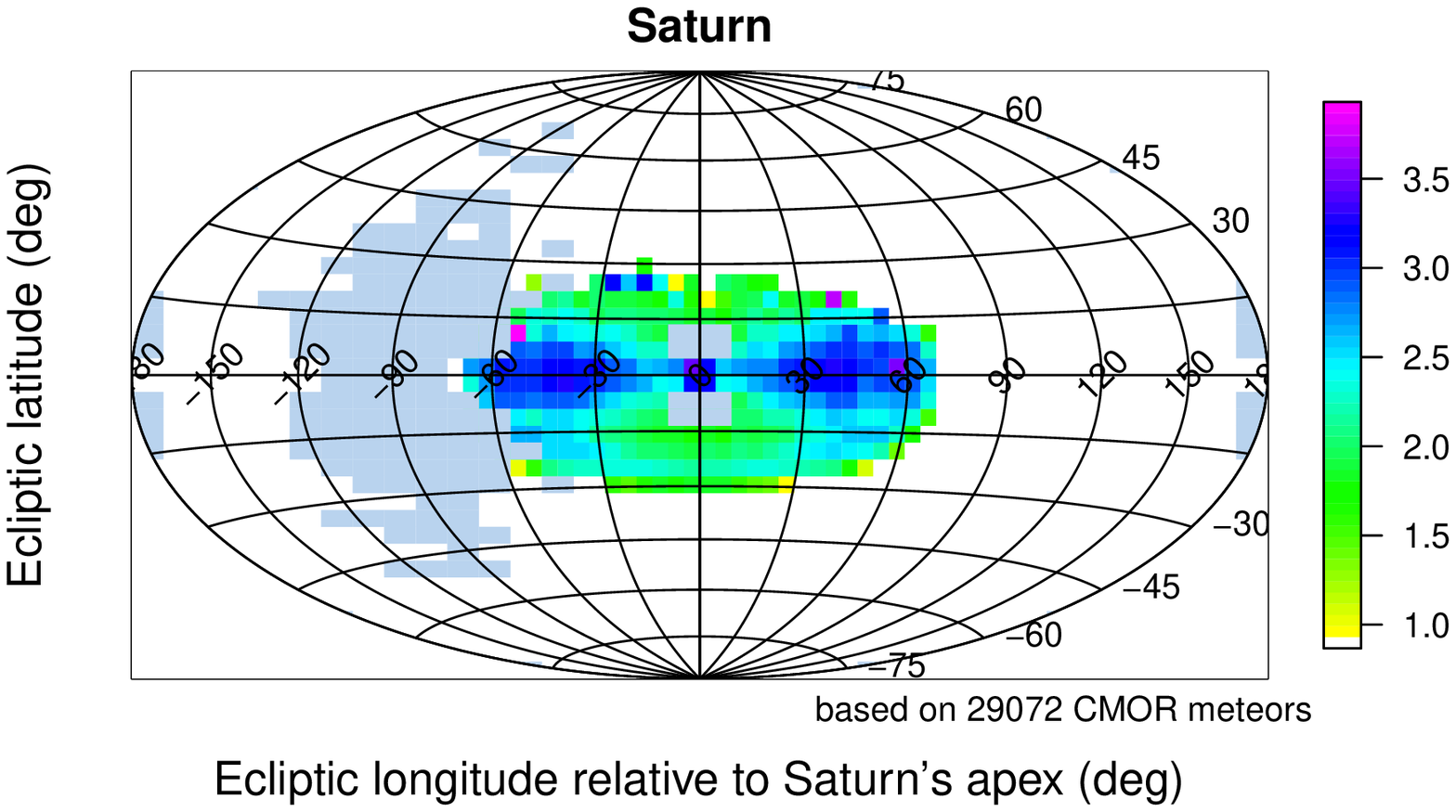}  &
        \includegraphics[width=12pc]{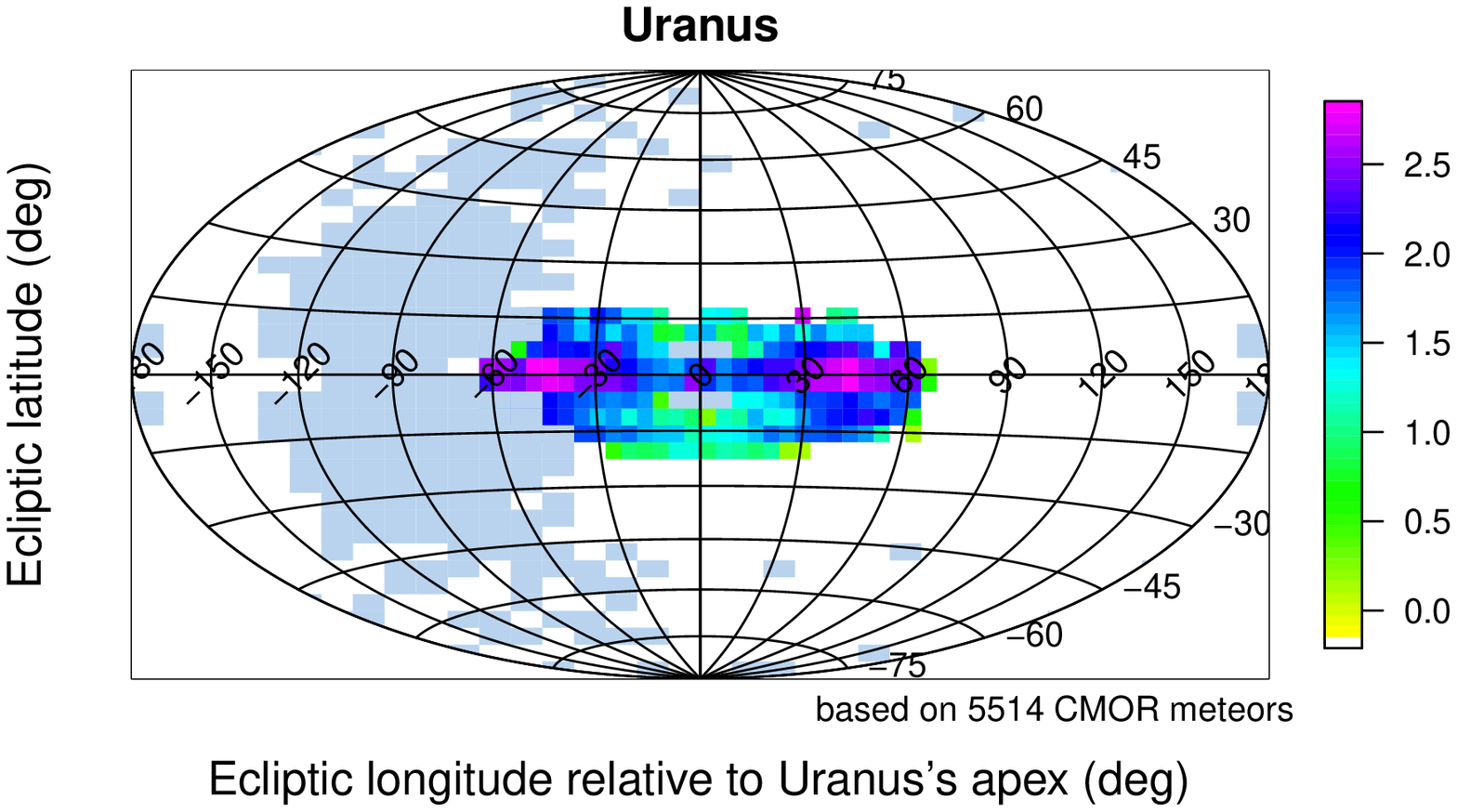}  &
        \includegraphics[width=12pc]{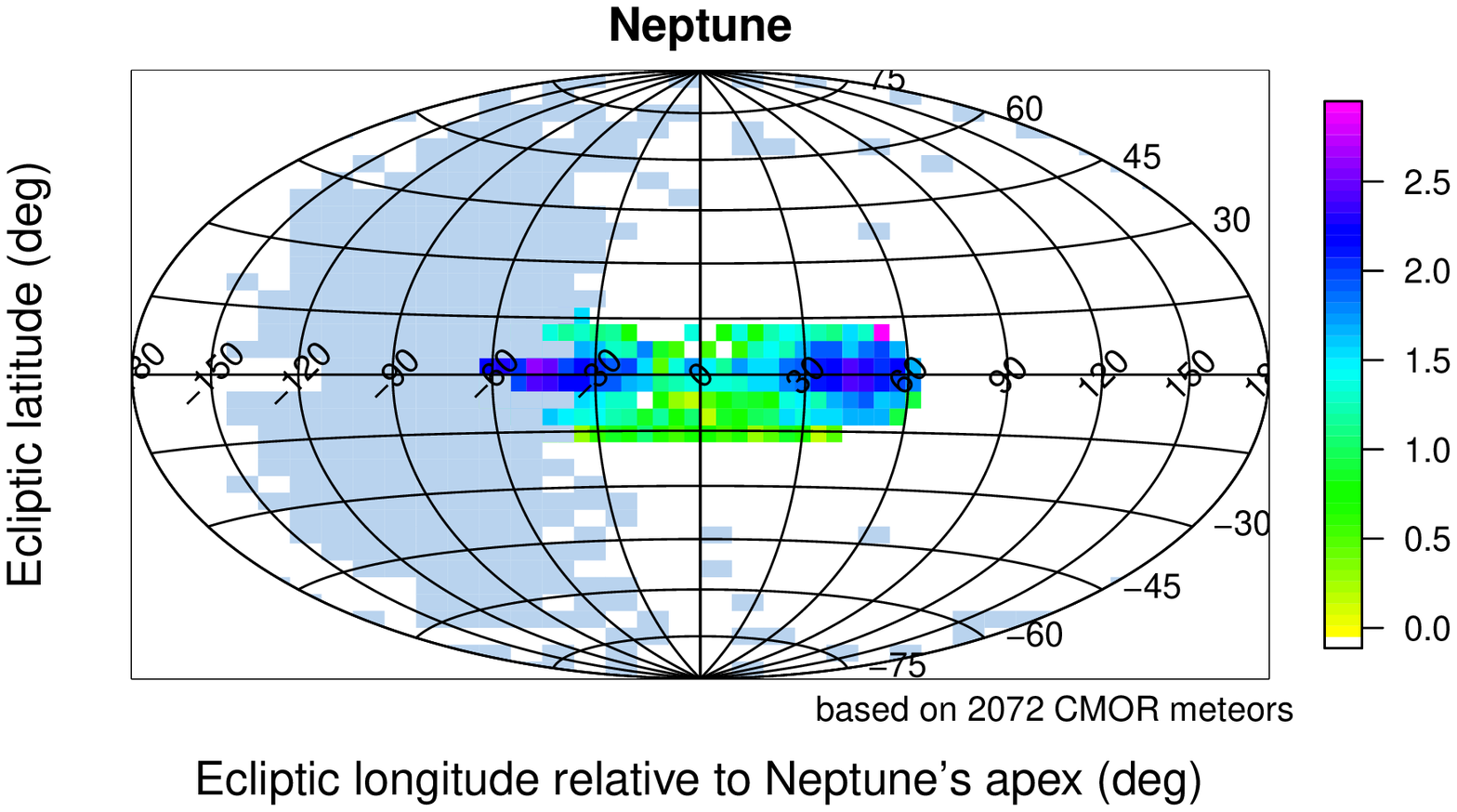}  
        
    \end{tabular}
    
    \caption{Radiant distribution of CMOR-detected meteors at Earth which pass within 0.01~AU of the terrestrial planets' orbits (1 AU for outer planets), in apex-centered ecliptic coordinates at the planet. The scale is the base-10 logarithm of the weighted number of    meteors at each radiant. The apex of the planet's way is in the center; the Sun is at $-90\degree$ longitude, and opposition at $+90\degree$. Pixels in grey indicate forbidden regions from which
   less than 1\% of possible speeds from those radiants can reach the
   Earth (i.e. the TF is less than 0.01); white pixels indicate
   radiants for which at least some meteors could reach the Earth
   (i.e. the TF is greater than 0.01) but none are observed by
   CMOR. \label{fi:PME_radiants}}
\end{figure}

To present our results, we again use Mars as our prototype. The
reconstructed radiant distribution at this planet is shown in the top panel of
Fig.~\ref{fi:PME_radiants}, from 38178 CMOR meteors collected from
2011-2014 (about $1~\%$ of the total) whose orbits, when traced backward, have MOIDs with 0.01~AU of Mars' orbit.

The most prominent features are the Martian helion and anti-helion sporadic sources,
the regions at zero latitude near $-90\degree$ and $+90\degree$ in
apex-centered longitude respectively.  The north and south toroidal
sporadic sources, at $\pm 60 \degree$ latitude and zero longitude on Earth are well-measured. In fact the south toroidal source is quite distinct in
the figure. The north toroidal source, which is comparable in strength
to the southern on Earth, is absent. However, this is likely a
selection effect due to the geometrical constraint of having a node
both at Mars and the Earth. CMOR is in Earth's northern hemisphere and
detects meteoroids which are mostly at their descending node at
Earth. These same particles are at their ascending node at Mars, and
hence strike it from the south.  Given the near mirror symmetry around
the ecliptic plane of the Earth's PME, it is likely that Mars does in
fact have north and south toroidal sources of comparable strength. The
northern toroidal source cannot be said to be weaker based solely on
these measurements, rather we would contend that CMOR's sensitivity to
the Mars northern toroidal source is much less. Clearly it would be
advantageous to combine meteor data sets obtained from Earth's north and
south hemispheres (e.g. \cite{pokjanbro17}) to provide a better picture.

\begin{figure}[ht!]
    \centering
    %\vspace*{-1cm}
    \hspace*{-0.45in}
    \includegraphics[height=20pc]{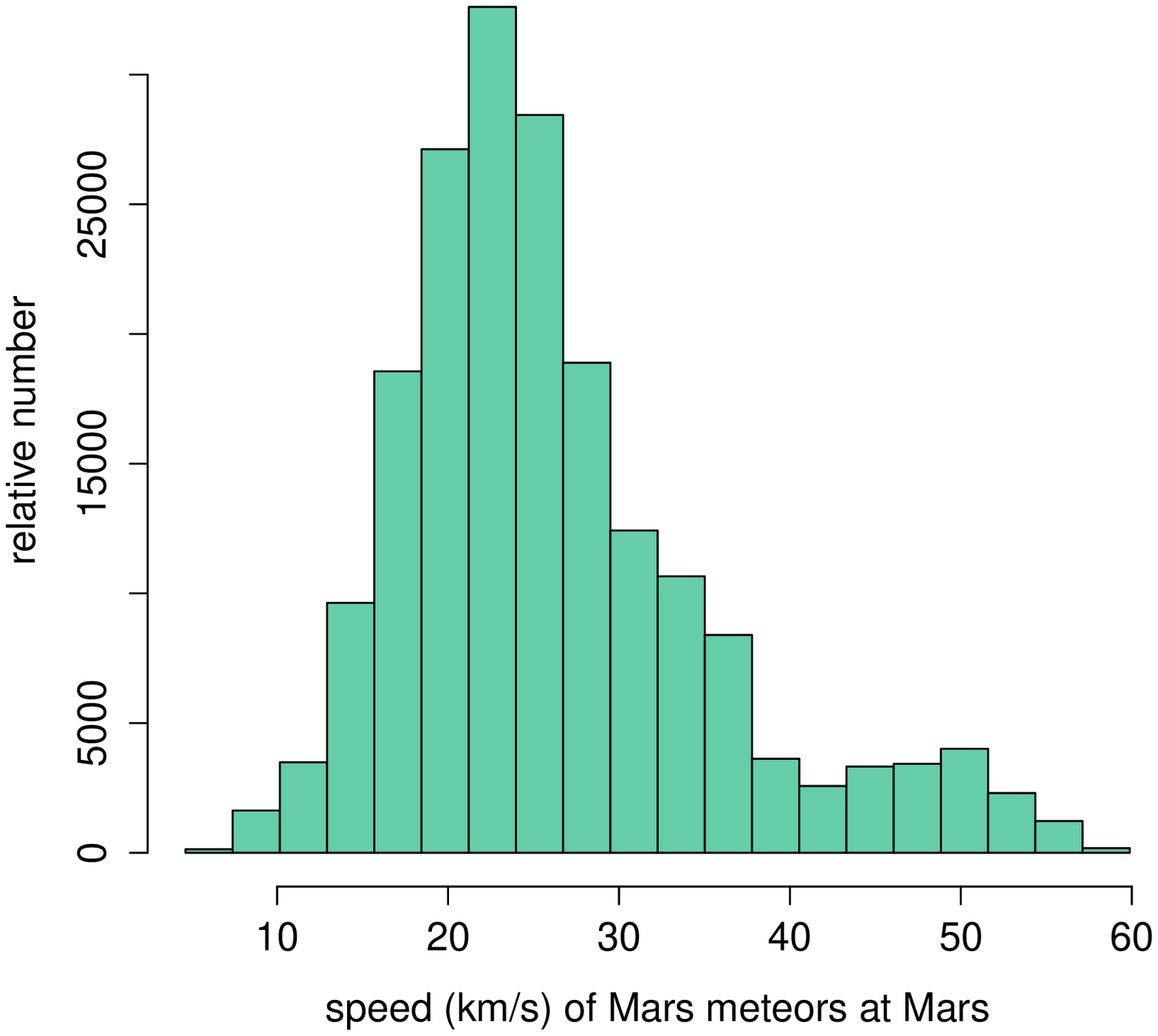}
    \begin{tabular}{ccc} % textwidth means that the image will be the width of the text
    \includegraphics[height=12pc]{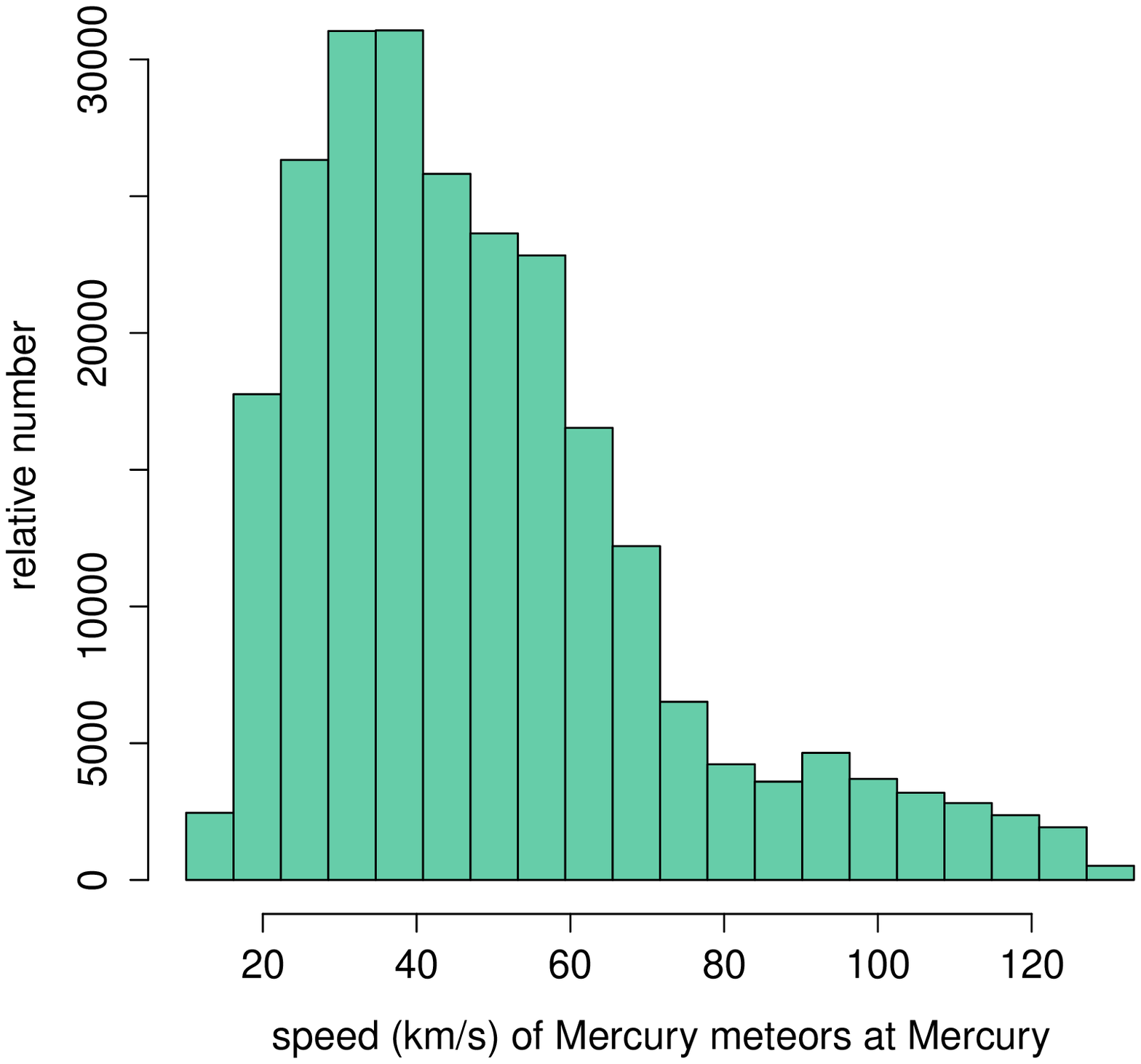} & 
    \includegraphics[height=12pc]{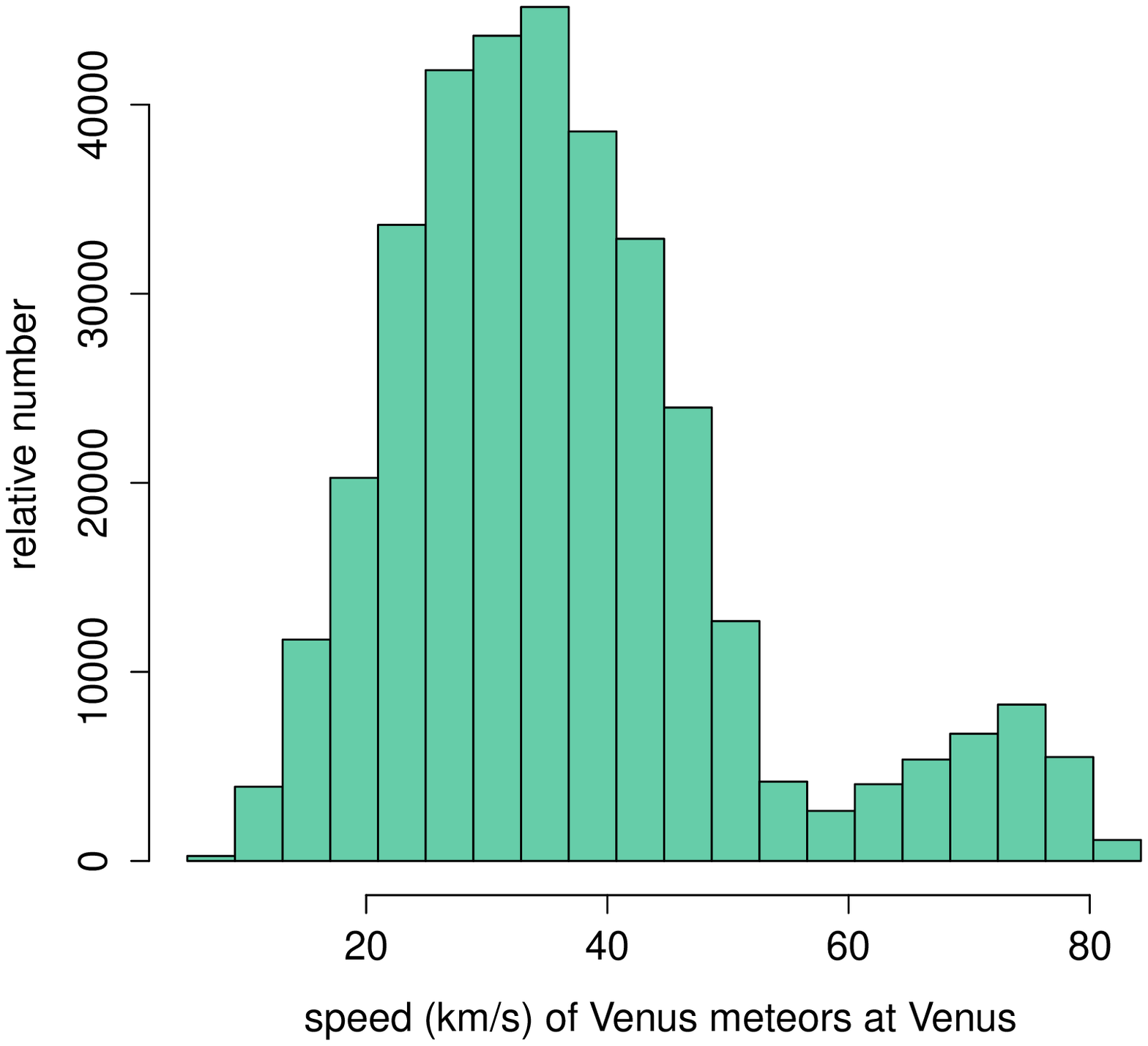} &
    \includegraphics[height=12pc]{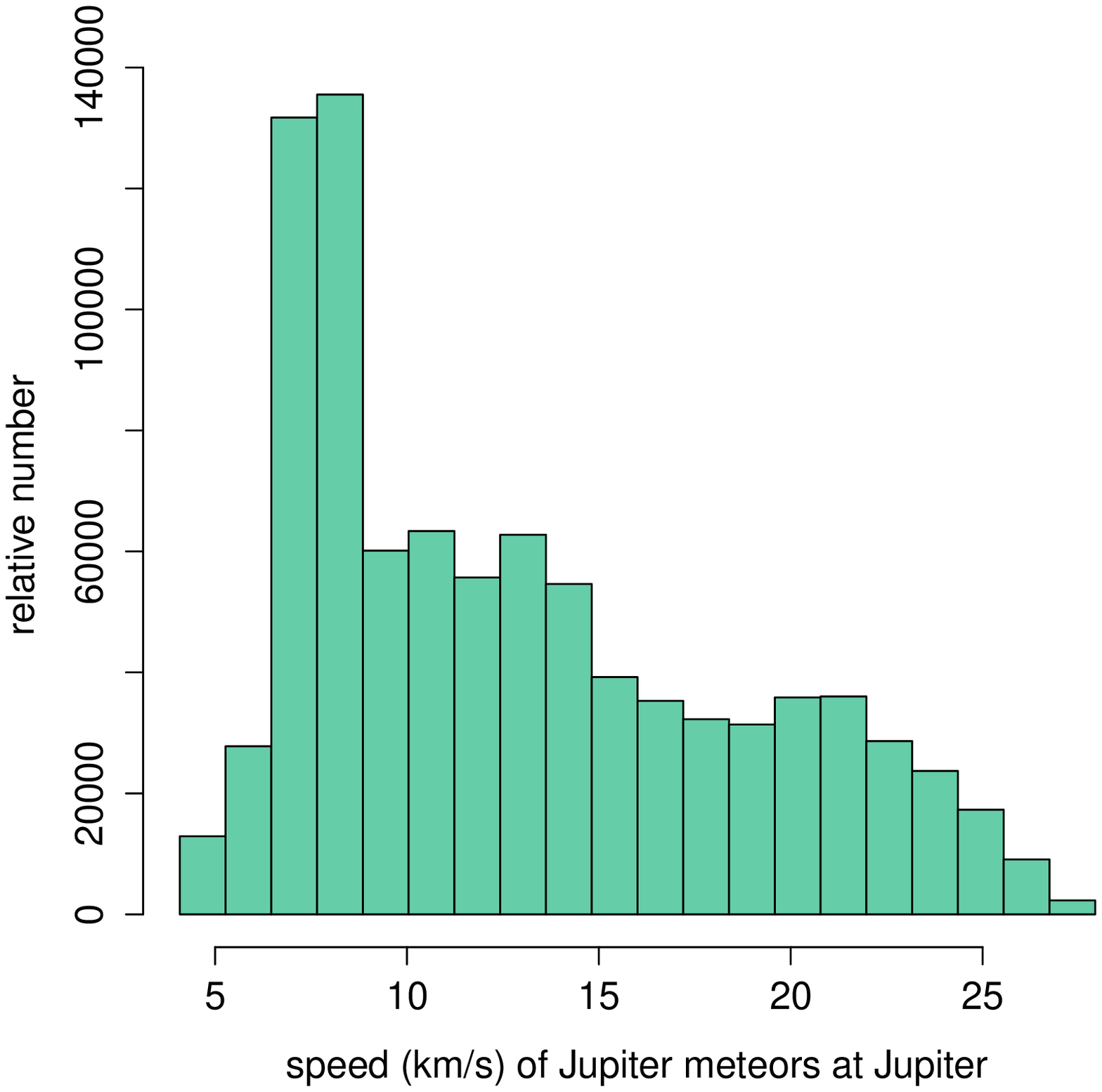} \\
    \includegraphics[height=12pc]{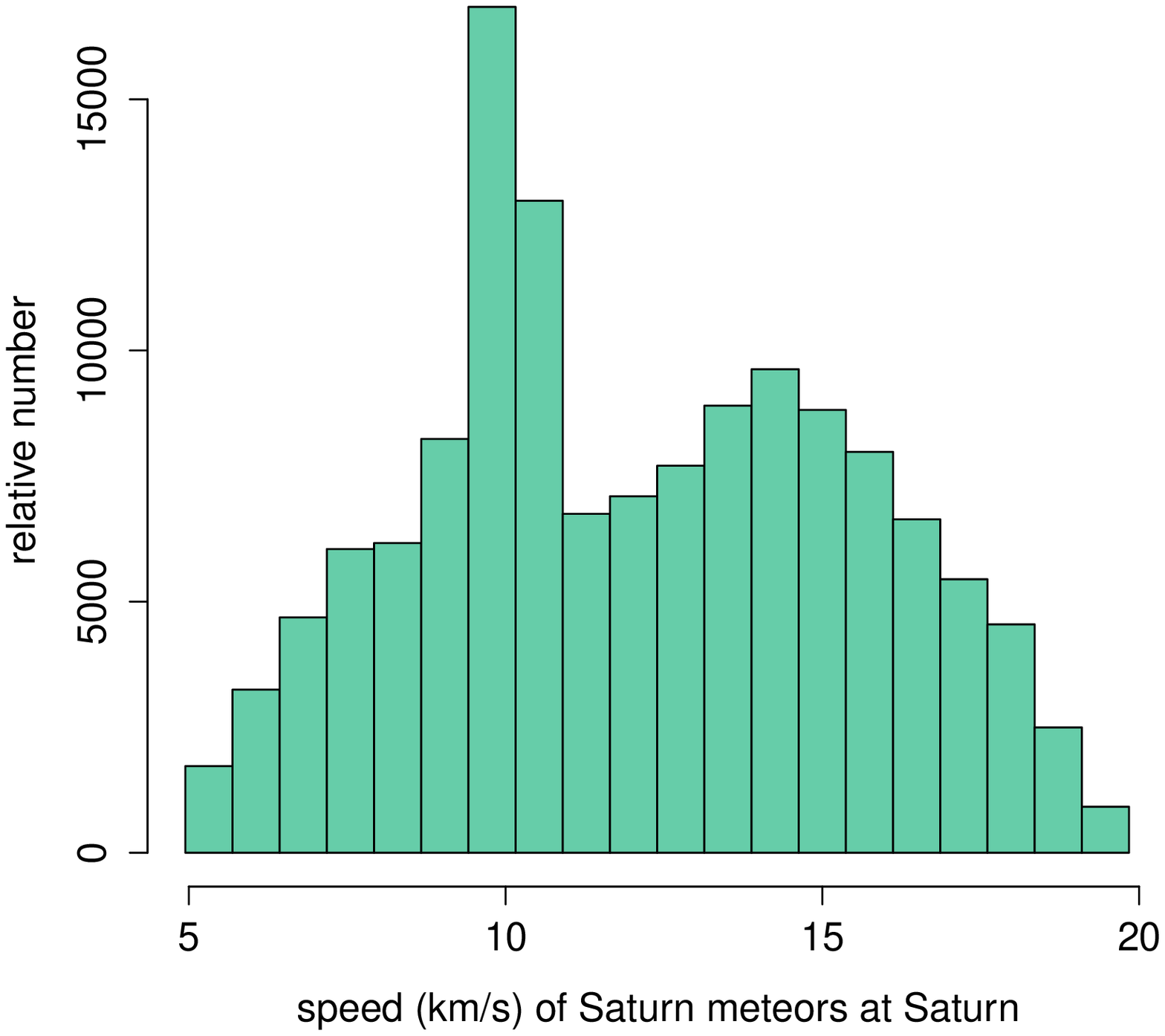}  &
    \includegraphics[height=12pc]{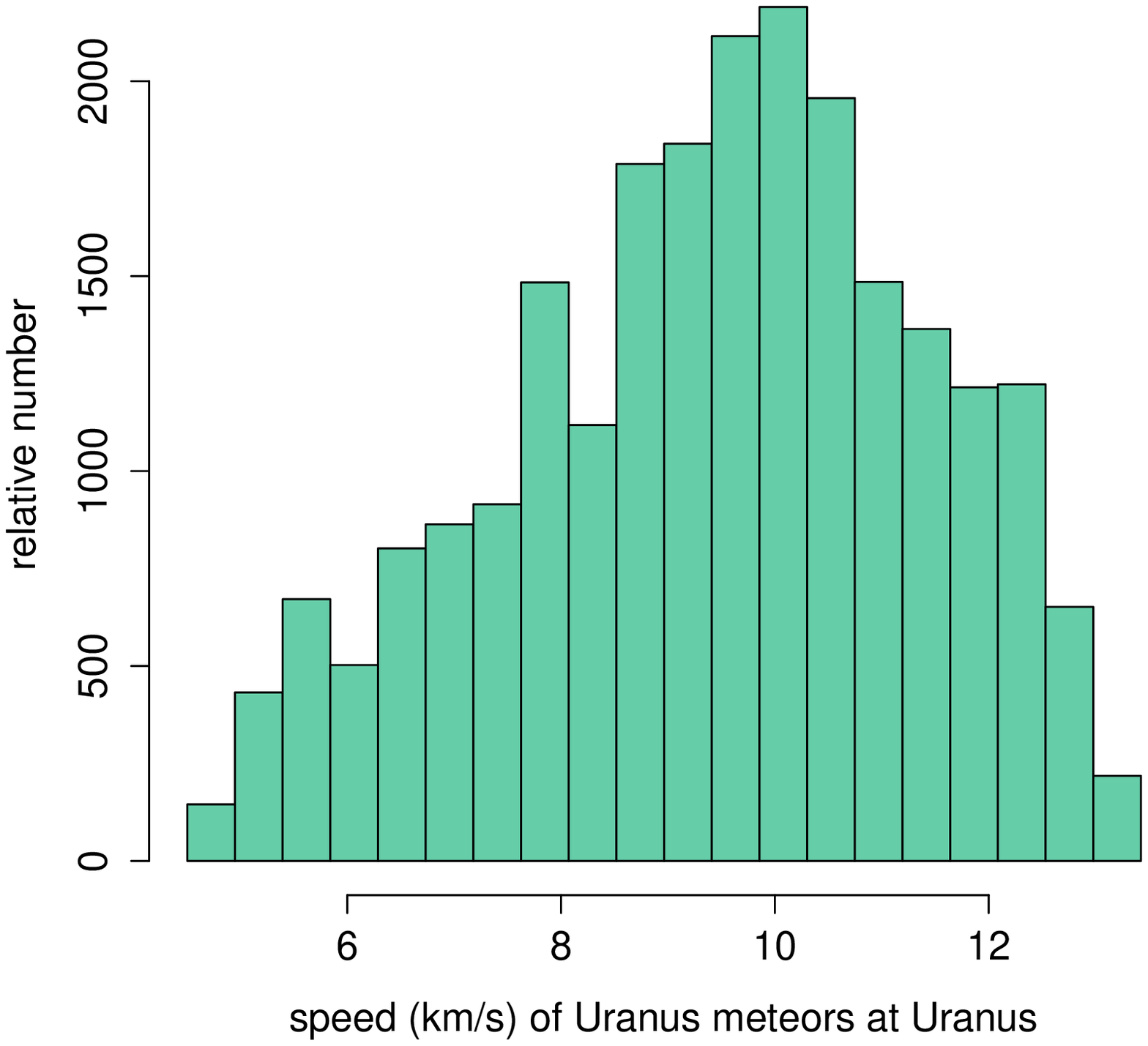}  & 
    \includegraphics[height=12pc]{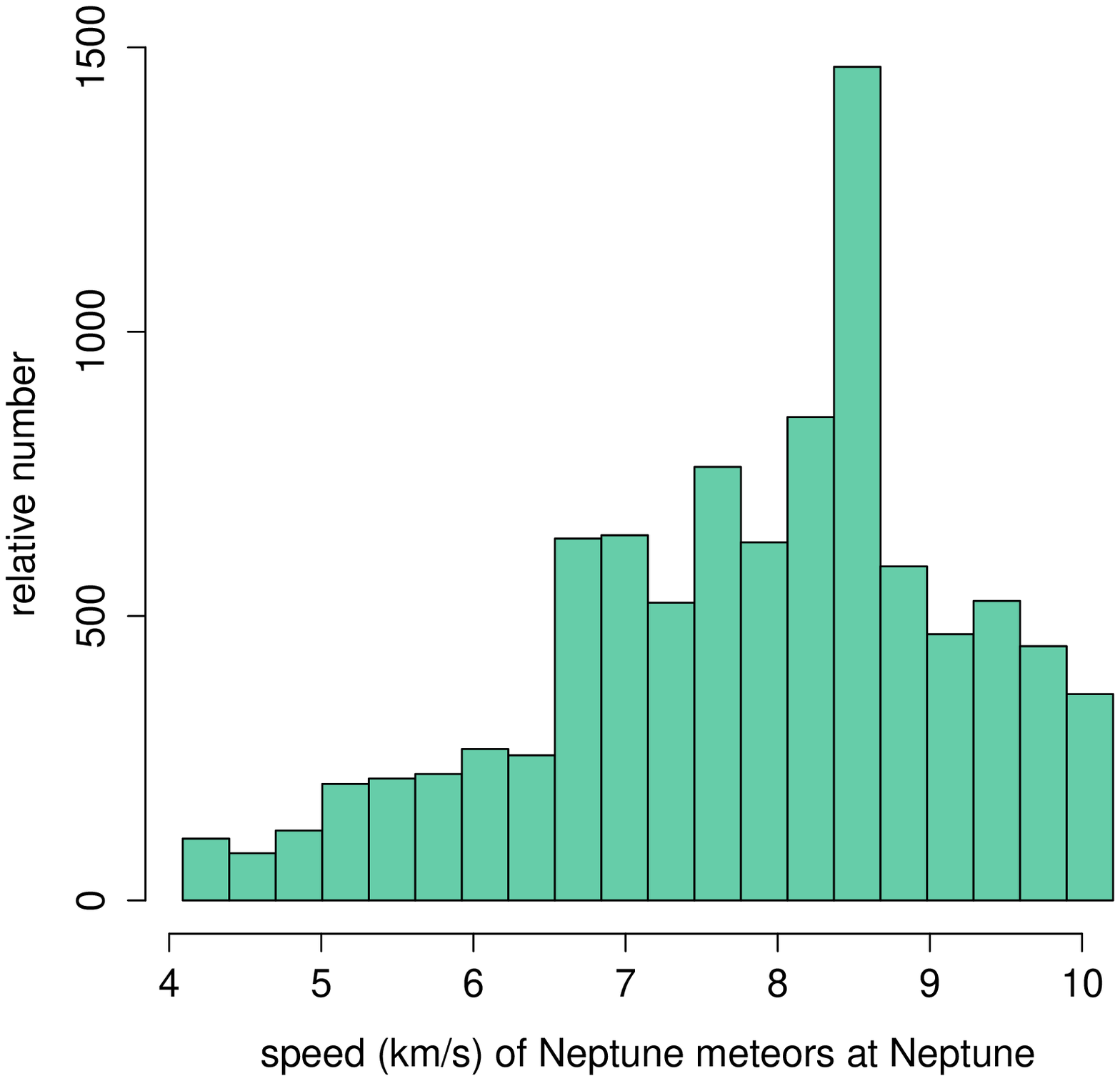}  
        
    \end{tabular}
    
    \caption{Meteoroid speed distribution, showing the speeds
      that would be measured at the planet, based on those CMOR meteors detected at Earth which intersect the planet's orbit. \label{fi:PME_speeds}}
\end{figure}

The apex sources near zero longitude and latitude are mostly
inaccessible, in a region where the TF is zero, but are sampled much
better along the ecliptic. The apex sources on Earth are associated
with retrograde meteors and hence with the highest speeds relative to
the planet. The top panel of Fig.~\ref{fi:PME_speeds} shows the speed distribution at
Mars. At the Earth, the overall speed distribution has two peaks, one
at 30 km/s and one at 55 km/s
\citep{tay95,satnaknis00,cloopphun02,jannolmei03,galbag04,hunoppclo04,chawoo04, chawoogal07,janclofen08,wievaucam09}. The higher speed peak is
associated with the apex sources. The fact that the distribution at
Mars shows two similar peaks indicates that our technique encompasses
a broad subset of the PME. In particular, the presence of the small
higher-speed peak in the top panel of Fig.~\ref{fi:PME_speeds} near 50 km/s at Mars
confirms that we can sample a subset of the high-speed Martian apex sporadic meteoroid population at Earth.

From these results, we predict with some confidence that Mars has
helion, anti-helion and south toroidal sporadic sources, and probably a
north toroidal source. The apex sources are not evident in the radiant
map, but the speed distribution suggests that they are being sampled,
albeit much more thinly, likely due to the low TF.

The meteoroid environment for the other planets , indeed for any
location within the Solar System, can be reconstructed in the same
manner. Here we present the results for the radiant distributions at the other planets in Figure~\ref{fi:PME_radiants}, and the speed distributions in Figure~\ref{fi:PME_speeds}. These are reconstructed from samples with sizes ranging from the lowest at 2072 meteors for Neptune to the largest at 156054 meteors for Jupiter, or 0.05\% to 3.6\% respectively of the 4.3 million meteors collected by CMOR from 2011 to 2014. We emphasize that this represents only part of the PME at each planet. However, the fact that there are radiants which are accessible through high TF at Earth from a given planet but not represented in CMOR measurements is an important constraint not previously known for different PMEs.

For the other terrestrial planets, the sporadic radiant features seen resemble those for Mars. There are prominent helion and anti-helion sporadic sources near $\pm90\degree$ from the apex, with a strong band stretching through the apex sources. The toroidal regions are less distinct but represented. The speed distributions show a clear bi-modality, again very similar to what is observed at Earth.

For the giant planets, we see meteors only in the region around the apex source, and there is a clear concentration of orbits in the ecliptic plane. Though portions of the helion and toroidal sources are notionally observable we don't see any meteors from these regions for any of the outer planets. However, we recall that the white regions in Fig.~\ref{fi:PME_radiants} are those where meteors at some speed from zero to the heliocentric escape speed could reach us, not necessarily those speeds which are actually populated at these planets. Thus the absence of toroidal or helion sources from these figures is not enough to conclude that there are no such meteors at the planet's themselves. They may exist but at speeds which result in orbits that do not reach us at Earth. In a similar vein, the density enhancements of radiants shown in Fig ~\ref{fi:PME_radiants} are subject to the same biases, and could be significantly affected by the absence from our sample of the meteor component which simply cannot reach the Earth. This geometrical constraint that we can only measure orbits passing near both the Earth and our planet means that the measurement of some components of PMEs will necessarily have to await {\it in situ} measurements.

\section{Conclusions}
Examination of the phase space in the meteoroid environments for each
of the planets reveals that a few to several percent of the meteoroids which have orbits passing close to other planetary orbits in the solar system can reach the Earth. From this we conclude that some measurement of
these environments is possible with Earth-based equipment, if we can
obtain a sufficiently large sample; and that calibration and
ground-truthing of model meteoroid environments is possible in principle, without the need to measure the planetary environment {\it in situ}.

A large sample of CMOR radar meteors observed at the Earth was then
examined to find those which are also part of the meteoroid environments
of other planets, and a partial meteoroid environment was reconstructed
for each of the planets.  Though only partial, it allows us to
predict features similar to the sporadic meteor sources observed at
Earth in the environments of Mars and the other terrestrial planets.
For the outer planets, high-speed retrograde meteors from their apex
sources are the best sampled. From CMOR observations of the meteoroid environment at Earth and our model, we predict a concentration of radiants in the ecliptic, but we have much sparser information on the existence of helion, anti-helion or toroidal sporadic sources. 

Thus the first measurements of the meteoroid environments of the planets have now been made.  In future works, through more careful corrections for the observational biases, larger CMOR datasets and higher fidelity modelling we believe the construction of well-calibrated models of other planetary meteoroid environments is possible.

\acknowledgments

PGB wishes to thank the Canada Research Chair program. This work was supported in part by the Natural Sciences and Engineering Research Council of Canada and NASA's Meteoroid Environment Office (MEO) through co-operative agreement NNX15AC94A.

\bibliography{Wiegert}
\newpage

% TABLES

%\documentstyle[aj_pt4]{article}
%\pagestyle{empty}
%\begin{document}
%\tablenum{1}

\begin{table}[p]
\caption{Percent of meteor radiants at a planet which reach Earth. Uniform refers to a random-on-the-sphere distribution of radiants at all speeds from zero to the local heliocentric escape speed. WCBV09 uses the radiant and speed distribution from the planetary meteoroid model of \cite{wievaucam09}.}
\centerline{
\begin{tabular}{cccccccc} \hline \hline
 Name    & Mercury & Venus & Mars & Jupiter & Saturn & Uranus & Neptune \\   \hline           
 Uniform (\%)& 2.8 &  5.2 &  9.0 & 6.2 & 5.5 & 4.3 & 3.5 \\ 
 WCBV09 (\%) & 8.7 & 23.7 & 14.4 & 2.0 & 1.3 & 1.3 & 2.6  
\end{tabular}}
\label{ta:TF}
\end{table}
%\end{document}

\end{document}